# Search of long-range forces of a neutron and atoms with a trap of ultracold neutrons

A.P. Serebrov[*], O.M. Zherebtsov, S.V. Sbitnev, V.E. Varlamov, A.V. Vassiljev, M.S. Lasakov

Petersburg Nuclear Physics Institute, Russian Academy of Sciences,
Gatchina, Leningrad District, 188300, Russia

A method of using a gravitational UCN spectrometer for search of long-range forces between neutrons and atoms is proposed. The constraints on the strength of long-range forces within the range of $10^{-10} - 10^{-4}$ cm can be obtained from experiments on measurements of the total cross section of interaction of UCN with atoms of noble gases (He, Ne, Ar, Kr ) and data on the coherent neutron scattering length of the nucleus. The first result of such an analysis is presented. Further prospects of UCN method are discussed.

## I. Introduction

Search of deviations of gravitational interaction from the $1/r^2$ law (inverse-square law) in the range of small distances is extremely important in order to verify both theories assuming existence of additional dimensions [1, 2] and super symmetric theories in which existence of new very light particles is assumed. The exchange of these particles leads to additional interactions between nucleons [3-5]. A review of theoretical and experimental works on search of deviations from the inverse-square law is presented in works [6, 7]. In the given work we will discuss forces which can appear at distances $10^{-10}$-$10^{-4}$ cm. From the point of view of search of deviations of gravitational interaction from the inverse-square law these forces should be defined as short-range forces. But in nuclear interactions a characteristic scale of distances is of the order of $10^{-13}$ cm, thereby for nuclear physics the interaction at distances of $10^{-10}$-$10^{-4}$ cm is carried on by long-range forces. We have stopped on the definition of long-range forces because it is a question of interaction of a neutron with a nucleus. Such a term is specified in the title of our article.

There are different methods of search of long-range forces in the interaction of elementary particles [6-8]. Within the range of $10^{-11}$-$10^{-9}$ cm investigations are carried out by means of neutrons at the energy of the order of electron volt [9, 10]. For distances $10^{-4}$-$10^{-2}$ cm the laboratory experiments on gravitational interactions of bodies are performed [11-20]. Within the range of $10^{-10}$ - $10^{-4}$ cm there are rather effective methods using thermal and cold neutrons [9, 21]. The present paper will discuss the possibility of using ultracold neutrons (UCN) for the range $10^{-10}$-$10^{-4}$ cm.

The scattering amplitude of a neutron by atoms can be expressed in the following way

$$f(q) = f_{nucl} + f_{n-e}(q) + f_{long\ range}(q), \qquad (1)$$

where $f_{nucl}$ is a nuclear scattering amplitude, which is usually expressed in terms of scattering length $b$, $f_{nucl} = -b$. $f_{n-e}(q)$ is the amplitude of neutron-electron scattering, which arises due to neutron scattering by charges distributed inside the nucleus and the electron shell of atoms. Further we will not consider contribution from $n-e$ interaction, because this effect occurs mainly for fast neutrons [22]. The following term in the equation (1) does relate to a hypothetical long-range interaction (compared to the nuclear one) of a neutron with a nucleus. $f_{long\ range}(q)$ is a spin independent amplitude of interaction which is likely to arise as a result of exchange by a scalar or vector boson. In case of a scalar type of interaction the potential of interaction is written as an attractive potential, for a vector boson exchange the potential of interaction is written as a repulsive one

$$\varphi(r) = \frac{\pm g_\pm^2 M \hbar c\, e^{-r/\lambda}}{4\pi r}, \qquad (2)$$

where $M$ is mass of interacting particle expressed in units of nucleon mass $m_n$, $\lambda$ is effective radius of interaction, $g_\pm^2$ is dimensionless coupling constant. It should be noted that in general case $M = m_1 m_2 / m_n^2$, where $m_1$, $m_2$ are masses of interacting particles. In formula (2) the upper sign corresponds to a vector type of interaction, whereas the lower sign corresponds to a scalar type of interaction.

---
[*] E-mail; serebrov@pnpi.spb.ru
Phone : +7 81371 46001
Fax : +7 81371 30072

In a similar way, the amplitude within the Born approximation can be presented in the form

$$f_{long\_range}(q) = -\frac{m}{2\pi\hbar^2}\int \varphi(r) e^{-i\vec{q}\vec{r}} dV = \mp \frac{2m}{\hbar^2}\frac{g_\pm^2 M \hbar c}{4\pi}\frac{\lambda^2}{(\lambda q)^2+1}, \qquad (3)$$

where $m$ is a reduced mass $m = \dfrac{m_n m_A}{m_n + m_A}$, mass of an atom of gas $m_A = m_n M$, $q = |\vec{k}' - \vec{k}|$ is a momentum transferred to a neutron, $\vec{k}$ and $\vec{k}'$ are wave vectors of the particle within the center of mass before and after collision. The momentum $q$ is bound with the neutron recoil energy $\varepsilon$ by a simple ratio: $q = \dfrac{\sqrt{2\varepsilon m_n}}{\hbar}$.

An experimental search of additional terms in the scattering amplitude can be based on the fact that a long-range interaction gives contribution to the scattering amplitude either at small transferred momentum $q$ or at small scattering angles. The scattering amplitude at $\theta = 0$ or $q = 0$ can be measured at high accuracy in neutron-optical experiments with an interferometer [21]. This result should be compared with $f_{nucl} = -b$ to find out presence of additional terms in equation (1). For example, comparison of measurements with interferometers and experiments with the Bragg diffractometer allows to obtain restrictions [9] which are strong enough.

Lately the method of studying quantum states of a neutron in the Earth gravitational field near the matter surface [23] has been actively discussed. However, there is lack of real statistics in these researches therefore we are going to propose more statistical method in this paper.

A direct method of research would be the method of a small angle scattering, as existence of long-range forces results in scattering occurring at small angles. In this method there are obvious problems caused by existence of a small angle scattering resulting from scattering on the texture of the matter and multiple scattering. Besides, initial divergence of a beam does not permit to distinguish scattering at very small angles from the beam divergence.

This paper suggests reconsidering an approach to the method of a small angle scattering and passing to registration of small recoil energy instead of small angles of scattering. A new approach suggests using gas of ultracold neutrons (UCN) as a target that collides with the flux of atoms being in the same trap. Criterion of a signal of scattering induced by long-range forces is a transfer of ultimately small recoil energy $\sim 10^{-7}$ eV, registered with a trap of ultracold neutrons.

For thermal neutrons the recoil energy to a neutron ~$10^{-7}$ eV corresponds to a scattering angle ~$2\cdot 10^{-3}$ radian, which is within divergence of an incident neutron beam. For cold neutrons this scattering angle is twice higher, but it does not yet exceed divergence of a neutron beam.

The method of an ultracold neutron trap filled with the investigated gas (He, Ne, Ar, Kr) allows recoil energy about $10^{-7}$ eV to be registered. On the other hand, the total cross section of UCN with gas will not involve areas with recoil energy smaller than $10^{-7}$ eV. Thus, it is possible to compare the scattering amplitude from interferometer measurements $f(0)$ with the amplitude obtained by the UCN method.

## 1. Experimental setup

One of the possible schemes of an experiment is shown in Fig. 1. It enables to use the available equipment of PNPI in ILL. UCN fill the trap (3) at an opened valve (2) and closed valves (5). The absorber (4) is placed in the bottom position at a distance "h" from the trap bottom. When equilibrium density in the trap is achieved, the valve (2) is closed. There are two foils installed at the trap entrance and the trap exit. The critical energy of foils is $E_{UCN}^{foil}=m_n gh$. Foils can be installed in UCN guide or removed from UCN guide. They will be used sometimes for forming UCN spectrum at the trap entrance and for analysis of UCN spectrum at the trap exit. Normally UCNs fill the trap when the entrance foil is removed. The UCNs are stored in the trap for predetermined time $t_{hold}$ to form the spectrum with maximal UCN energy equal to $m_n gh$. Then the absorber (4) is pulled up to the upper position near the top of the trap.

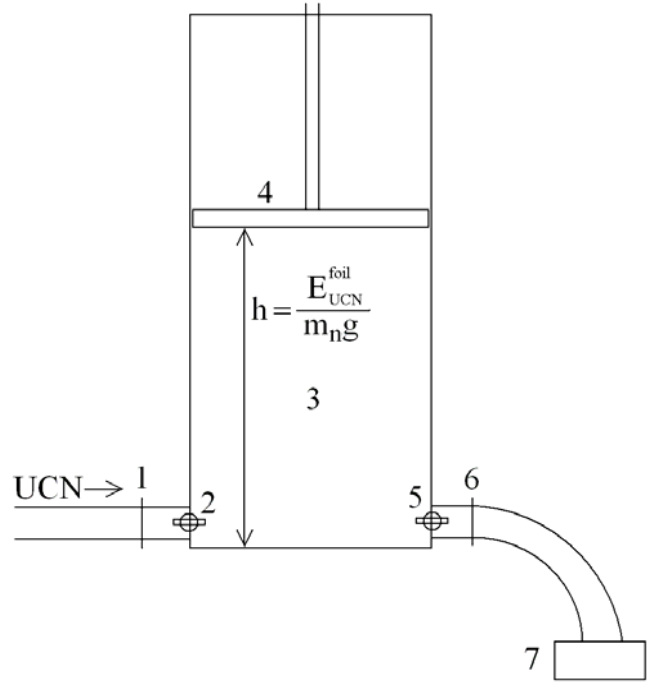

Fig1. The experiment setup. 1 – entrance foil with critical energy $E_{UCN}^{foil}=m_n gh$, 2 – the entrance valve, 3 – UCN trap with the critical energy $E_{UCN}^{trap}$, 4 – the absorber for formation of UCN spectrum, 5 – the exit valve, 6 – exit foil with critical energy $E_{UCN}^{foil}=m_n gh$, 7 – the UCN detector.

Ultracold neutrons interact with trap walls and with the investigated gas, which fills the trap. The trap and the investigated gas are maintained at room temperature. The temperature of UCN gas is $10^{-3}$ K. In coherent reflection from matter the energy of UCN is conserved. Inelastic scattering occurs when a neutron is scattered by atoms of gas and when UCN penetrate into the substance in reflecting from the wall. In both cases there is energy transfer by the order of $kT$. The energy transfer of the order of $10^{-7}$ eV is of low probability for the above mentioned processes. Nevertheless in reflecting from the substance there is a quasi-elastic scattering which was revealed experimentally [24]. Due to long range interaction with atoms of gas the quasi-elastic scattering at the recoil energy of the order $10^{-7}$ eV would be also possible. Processes of the energy transfer of the order of $kT$ and $\sim 10^{-7}$ eV are easily distinguished in the present installation, as UCN having obtained energy $\sim kT$ leave an experimental trap. Such neutrons are not detected. Neutrons which have obtained a small recoil energy can still be stored in the trap, if their energy near the bottom is less than critical energy of the trap. Critical energy of the foil is equal to $m_n g h$. Therefore neutrons which have obtained a small recoil energy can overcome a potential barrier of the foil (6) and finally can be registered by a detector (7). For registration of these neutrons the valve (5) is opened just after lifting the absorber. The closed valve (5) is used for measuring the background at the detector (7). To distinguish between processes of quasi-elastic scattering on the surface of the trap and scattering with the investigated gas, measurements are to be made both with the investigated gas and without it. The above described scheme of measurement permits to determine an extremely low energy transfer. Such a scheme was used in our measurements of lower energy up scattering of UCN from the trap walls [24].

For measuring the total cross section of UCN interaction with gas the detector (7) is applied to measure the number of UCN in the trap after different holding time with closed valve (5). In this case the foil (6) is removed from the guide. Measuring UCN storage time at different gas pressure, we can determine the total cross section of UCN interaction with atoms. The measurement of UCN storage time in the trap $\tau_{stor}$ is a conventional procedure. It consist from measurements of the number of UCN in the trap $N(t_1)$ at the moment $t_1$ after closing the entrance valve (2) and the number of UCN in the trap $N(t_2)$ at the moment $t_2$. The storage time $\tau_{stor}$ is determined according to the formula $\tau_{stor} = \ln(N(t_1)/N(t_2))/(t_2 - t_1)$. As UCN are sensitive to a small energy transfer, this cross section will include interaction induced by long-range forces. We can compare the obtained result with nuclear scattering cross section.

Summing up, one can conclude that two experimental methods have been discussed: the method of total cross-section measurements where the exit foil (6) is not used and the method of above-barrier neutron measurements using the exit foil (6). Potential sensitivity of both methods will be discussed below.

## 2. Numerical calculations and estimations

Let us calculate the differential cross section depending on the recoil energy transferred to the ultracold neutron. To simplify the problem we will assume UCN before collision to be at rest. The atoms of gas are scattered at UCN. Neutrons obtain the recoil energy. We will consider the amplitude of an additional long-range interaction of a neutron with an atom consisting of $M$ nucleons.

The differential cross section of scattering of a neutron with an atom should take into account the amplitude of nuclear scattering and that of scattering (3) due to an additional contribution from potential (2):

$$d\sigma = \left| f_{nucl} + f_{long\ range} \right|^2 d\Omega = \left( b_{free\_nucl}^2 \pm \frac{g_\pm^2 M}{\pi} \frac{b_{free\_nucl} mc^2}{\hbar c} \frac{\lambda^2}{(\lambda q)^2 + 1} + f_{long\ range}^2 \right) d\Omega, \quad (4)$$

where the element of solid angle $d\Omega$ is bound with energy of an incident atom - $E_A$ according to the following formula:

$$d\Omega = \frac{\pi (M+1)^2}{M} \frac{d\varepsilon}{E_A} \quad (5)$$

For the differential cross section the following expression has been derived

$$d\sigma = \pi \left( \frac{(M+1)^2}{M} b_{free\_nucl}^2 \pm \frac{g_\pm^2 M (M+1)}{\pi} \frac{b_{free\_nucl} m_n c^2}{\hbar c} \frac{\lambda^2}{2m_n \varepsilon \lambda^2 / \hbar^2 + 1} + \right.$$

$$\left. + \frac{g_\pm^4 M^3 \lambda^4}{4\pi^2} \left( \frac{m_n c}{\hbar} \right)^2 \Big/ \left( 2m_n \varepsilon \lambda^2 / \hbar^2 + 1 \right)^2 \right) \frac{d\varepsilon}{E_A}. \quad (6)$$

Fig. 2 (a), (b) shows the dependence of the differential cross section of scattering on the recoil energy for two different cases of $\lambda = 10^{-8}$ cm and $\lambda = 10^{-6}$ cm for a repulsive and an attractive potential, correspondently. The given calculations have been made in Fig. 2 for the fixed energy of an incident atom of helium equal to 2.5 $10^{-2}$ eV. For $\lambda = 10^{-8}$ the differential scattering cross section spans the full range of energies from 0 to $\varepsilon_{max} = E_A 4M/(1+M)^2$. At $\lambda = 10^{-6}$ cm, the differential cross section changes rapidly at recoil energies of the order of $10^{-7}$ eV. The proposed method is most sensitive for $\lambda = 10^{-6}$ cm. At $\lambda = 10^{-4}$ cm recoil energies are too small.

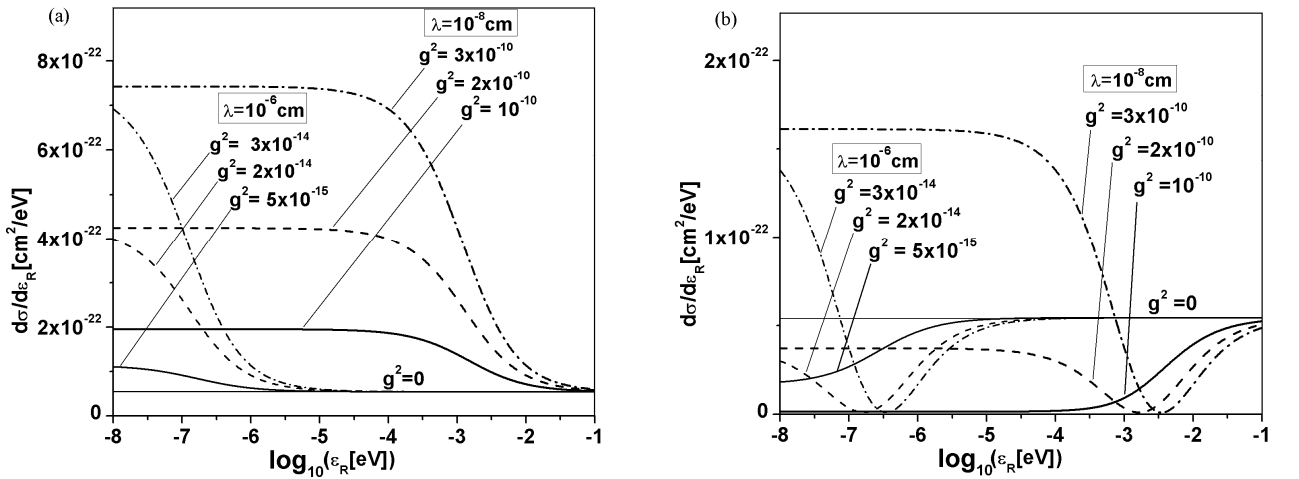

Fig. 2 Dependence of the differential cross section on recoil energy $\varepsilon_R$ transferred to a neutron for various values of parameter λ. (a) the case of a repulsive potential, (b) the case of an attractive potential.

Now we will integrate expression (6) over recoil energy from $\varepsilon_1$ up to $\varepsilon_2$.

$$\sigma(\varepsilon_2, \varepsilon_1, E_A) = \pi \left[ \frac{(M+1)^2}{M} b_{free\_nucl}^2 (\varepsilon_2 - \varepsilon_1) \pm g_\pm^2 M \frac{\hbar c\ b_{free\_nucl}(M+1)}{2\pi} \ln \left( \frac{2m_n \lambda^2 \varepsilon_2 / \hbar^2 + 1}{2m_n \lambda^2 \varepsilon_1 / \hbar^2 + 1} \right) + \right.$$

$$+\left(g_{\pm}^{2}M\right)^{2}\left(\frac{m_{n}c}{2\pi\hbar}\right)^{2}\frac{M\lambda^{4}(\varepsilon_{2}-\varepsilon_{1})}{\left(2m_{n}\lambda^{2}\varepsilon_{2}/\hbar^{2}+1\right)\left(2m_{n}\lambda^{2}\varepsilon_{1}/\hbar^{2}+1\right)}\Bigg]\frac{1}{E_{A}}. \tag{7}$$

The integral UCN scattering cross section when UCN escape from the trap with critical trap energy ($E_{UCN}^{trap}$) is:

$$\sigma^{escape}(E_{UCN}^{trap},E_{A})=\pi\Bigg[4b_{free\_nucl}^{2}\left(1-\frac{E_{UCN}^{trap}(M+1)^{2}}{4E_{A}M}\right)\pm g_{\pm}^{2}M\frac{\hbar c\, b_{free\_nucl}(M+1)}{2\pi E_{A}}\ln\left(\frac{8m_{n}\lambda^{2}E_{A}M/\hbar^{2}(M+1)^{2}+1}{2m_{n}\lambda^{2}E_{UCN}^{trap}/\hbar^{2}+1}\right)+$$

$$+\left(g_{\pm}^{2}M\right)^{2}\left(\frac{m_{n}cM}{\pi\hbar(M+1)}\right)^{2}\frac{\lambda^{4}\left(1-E_{UCN}^{trap}(M+1)^{2}/4E_{A}M\right)}{\left(8m_{n}M\lambda^{2}E_{A}/(M+1)\hbar^{2}+1\right)\left(2m_{n}\lambda^{2}E_{UCN}^{trap}/\hbar^{2}+1\right)}\Bigg]. \tag{8}$$

The lower energy UCN scattering cross section when UCN are still stored in the trap is:

$$\sigma^{low}(E_{UCN}^{trap},E_{A})=\pi\Bigg[\frac{(M+1)^{2}}{M}b_{free\_nucl}^{2}\frac{E_{UCN}^{trap}}{E_{A}}\pm g_{\pm}^{2}M\frac{\hbar c\, b_{free\_nucl}(M+1)}{2\pi E_{A}}\ln\left(2m_{n}\lambda^{2}E_{UCN}^{trap}/\hbar^{2}+1\right)+$$

$$+\left(g_{\pm}^{2}M\right)^{2}\left(\frac{m_{n}c}{2\pi\hbar}\right)^{2}\frac{M\lambda^{4}E_{UCN}^{trap}/E_{A}}{\left(2m_{n}\lambda^{2}E_{UCN}^{trap}/\hbar^{2}+1\right)}\Bigg]. \tag{9}$$

The total scattering cross section of UCN with an atom is:

$$\sigma_{scatt}^{total}(E_{A})=\pi\Bigg[4b_{free\_nucl}^{2}\pm g_{\pm}^{2}M\frac{\hbar c\, b_{free\_nucl}(M+1)}{2\pi E_{A}}\ln\left(8m_{n}\lambda^{2}E_{A}M/\hbar^{2}(M+1)^{2}+1\right)+$$

$$+\left(g_{\pm}^{2}M\right)^{2}\left(\frac{m_{n}cM}{\pi\hbar(M+1)}\right)^{2}\frac{\lambda^{4}}{\left(8m_{n}M\lambda^{2}E_{A}/(M+1)\hbar^{2}+1\right)}\Bigg]. \tag{10}$$

It should be mentioned that for the sake of simplification we have assumed initial UCN energy to be equal to zero. Such a simplification does not matter but it makes the calculation much easier. Formulae (7-10) are written for the fixed kinetic energy of an atom. For further calculations we should integrate over the flux of incident atoms. As it has been noted above, the installation setup enables to measure the total cross sections of interaction of UCN with gas using the detector without exit foil (6), and the differential cross sections of a very small energy transfer using the detector with exit foil (6). In the following paragraph the first experimental observables are considered in detail.

## 3. Measuring the total interaction cross section of a neutron and atoms of gas

The probability of UCN storage in a trap is the sum of probability of UCN losses:
$$\tau_{stor}^{-1\,total}=\tau_{n}^{-1}+\tau_{stor}^{-1\,gas}+\tau_{stor}^{-1\,walls}, \tag{11}$$

where $\tau_n^{-1}$ is the neutron decay probability, $\tau_{stor}^{-1\ gas}$ is the probability of UCN losses due to interaction with gas atoms and $\tau_{stor}^{-1\ walls}$ is the probability of UCN losses due to interaction with the trap walls.

The probability of UCN losses caused by neutron interaction with gas atoms can be measured as the difference of UCN storage probability in a trap with gas density $n_A$ and with zero gas density:

$$\tau_{stor}^{-1\ gas}(n_A) = \tau_{stor}^{-1\ total}(n_A) - \tau_{stor}^{-1\ total}(n_A = 0). \tag{12}$$

Now let us calculate the magnitude of value $\tau_{stor}^{-1\ gas}(n_A)$ taking into account an additional contribution made by the long-range interaction. The probability of UCN losses induced by collision with gas atoms can be written as follows:

$$\tau_{stor}^{-1\ gas}(n_A) = \int_{E_{min}}^{\infty} d\Phi(E_A) \int_{\varepsilon_{min}}^{E_A \frac{4M}{(M+1)^2}} d\sigma(\varepsilon) + n_A V_{2200} \sigma_{capt}^0 = n_A \bar{V}_A \sigma_A^{total}, \tag{13}$$

where $d\Phi(E_A)/dE_A$ is a flux of atoms incident on an ultra cold neutron, $d\sigma/d\varepsilon(\varepsilon)$ is a differential cross section depending on the recoil energy, according to formula (6), $\sigma_{capt}^0$ is the capture cross section reduced to neutron velocity $V_{2200} = 2.2 \times 10^5$ cm s$^{-1}$, $\sigma_A^{total}$ is the total cross section which consists of scattering cross section ($\sigma_{scat}$) and capture cross section ($\sigma_{capt}^0$): $\sigma_A^{total} = \sigma_{scat} + \sigma_{capt}^0 V_{2200} / \bar{V}_A$. (The scattering cross section ($\sigma_{scat}$) takes account of a nuclear and long-range interaction). $E_{min}$ is minimum energy of atoms after colliding with which a neutron is able to escape the trap $E_{min} = E_{UCN}^{trap}(M+1)^2/4M$, $\varepsilon_{min}$ is minimum recoil energy when UCN escape from the trap. To simplify, the UCN initial energy is equal to zero $\varepsilon_{min} = E_{UCN}^{trap}$, $E_A 4M/(M+1)^2$ is maximum neutron recoil energy. The flux of atoms is

$$d\Phi(E_A)/dE_A = \frac{n_A \bar{V}_A}{(kT)^2} E_A \exp\left\{-\frac{E_A}{kT}\right\} \tag{14}$$

where $n_A$ is the number of atoms in cm$^{-3}$ at temperature $T$, $\bar{V}_A$ is the average velocity of atoms of mass $m_n M$ at temperature $T$, $\bar{V}_A = 4(kT/2\pi m_n M)^{1/2}$. The gas density $n_A[\text{cm}^{-3}] = 2.687 \cdot 10^{16} \times P_A[\text{mbar}] \times 273/T[\text{K}]$, where $P_A$ is an experimentally measured gas pressure.

We can rewrite formula (13) in the following way:

$$(\tau_{stor}^{gas} n_A \bar{V}_A)^{-1} - \sigma_{capt}^0 \bar{V}_{th}/\bar{V}_A = \frac{\pi(M+1)^2}{M} b_{free\_nucl}^2 \int_{E_{UCN}^{trap} \frac{(M+1)^2}{4M}}^{\infty} dE_A \int_{E_{UCN}^{trap}}^{E_A \frac{4M}{(M+1)^2}} \frac{E_A}{(kT)^2} e^{-E_A/kT} \times$$

$$\times \left[ 1 \pm \frac{g_\pm^2 M^2}{\pi(M+1)} \frac{m_n c^2}{b_{free\_nucl} \hbar c} \frac{\lambda^2}{2m_n \varepsilon \lambda^2/\hbar^2 + 1} + \right.$$

$$\left. + (g_\pm^2 M)^2 \frac{M^2}{(M+1)^2} \left(\frac{m_n c}{2\pi \hbar b_{free\_nucl}}\right)^2 \frac{\lambda^4}{(2m_n \lambda^2 \varepsilon/\hbar^2 + 1)^2} \right] \frac{d\varepsilon}{E_A}. \tag{15}$$

After integration one can subtract contribution made by long-range forces as follows:

$$\Delta_1^{(P\tau)-nucl} = \left[ \frac{(\tau_{stor}^{gas} n_A \overline{V}_A)^{-1} - \sigma_{capt}^0 V_{2200}/\overline{V}_A}{4\pi b_{free\_nucl}^2} e^{\frac{E_{UCN}^{trap}}{kT}\frac{(M+1)^2}{4M}} - 1 \right] = \pm \frac{g_\pm^2 M(M+1)}{8\pi b_{free\_nucl}} \left(\frac{\hbar c}{kT}\right) e^{z(\lambda)} E_1(z(\lambda)) +$$

$$+ \frac{g_\pm^4}{4} \frac{M^2(M+1)^2}{(8\pi b_{free\_nucl})^2} \left(\frac{\hbar c}{kT}\right)^2 \frac{e^{z(\lambda)} E_2(z(\lambda))}{z}, \qquad (16)$$

where $E_1(z)$ and $E_2(z)$ are exponential integrals. The value $z(\lambda) \equiv \frac{(M+1)^2}{4M}\left(\frac{\hbar^2}{2m_n kT\lambda^2} + \frac{E_{UCN}^{trap}}{kT}\right)$ is a function of a few variables: $M$, $\lambda$ and $T$.

The expression in the squared brackets (on the left side of formula (16)) is an expected experimental effect resulting from the long-range interaction. It is defined as $\Delta_1^{(P\tau)-nucl}$ since in this analysis we compare the nuclear scattering cross section and the scattering cross section obtained from $(P\tau)$- measurements with UCN. We assume that information on the nuclear scattering cross section ($4\pi b_{free\_nucl}^2$) is available. In fact this information can be obtained from a neutron scattering experiment at neutron energy about 1eV.

Let us calculate the value of an expected experimental effect $\Delta_1^{(P\tau)-nucl}$ depending on values $g^2$ and $\lambda$. The case of a repulsive potential (a vector boson) and the case of an attractive potential (a scalar boson) significantly differ by the effect shape. In the case of a vector boson the effect is positive for any values $g^2$ and $\lambda$. For a scalar boson the effect can change a sign depending on values $g^2$ and $\lambda$. The shape of a possible effect for both cases is shown in Fig. 3 (a), (c). Fig. 3 (b), (d) shows the correlation between $g^2$ and $\lambda$, which arises when this surface is crossed by the planes $\Delta_1^{(P\tau)-nucl} = \pm 0.3$, $\Delta_1^{(P\tau)-nucl} = \pm 0.03$ and $\Delta_1^{(P\tau)-nucl} = \pm 0.003$. In case when $\varphi > 0$, the value $\Delta_1^{(P\tau)-nucl}$ can be only positive. In case when $\varphi < 0$, $\Delta_1^{(P\tau)-nucl}$ can have any sign. Therefore at $\Delta_1^{(P\tau)-nucl} > 0$ the determination of a potential sign from an experiment is ambiguous. As a rule we are compelled to analyze both cases: with $\varphi > 0$ and with $\varphi < 0$.

As seen from Fig. 3 the method of comparing the scattering cross section of UCN and nuclear scattering cross section becomes insensitive in the area of $10^{-8} - 10^{-4}$cm. This is due to a slightly increasing logarithmic dependence in (8) at sufficiently high energy of incident atoms at room temperature (lowering the temperature of gas may result in some progress). Thus, the integral measurement method of UCN essentially determines the scattering cross section in the field of forces less than $10^{-8}$cm ($\lambda < 10^{-8}$cm). The method of measuring low-energy neutrons is sensitive down to $\lambda \approx \lambdabar_{UCN}$ i.e. $10^{-6}$cm $\left(\lambdabar_{UCN}^2 = \hbar^2/2m_n E_{UCN}^{trap}\right)$. (In formula (9) under the logarithm there is the squared ratio of $\lambda$ to $\lambdabar_{UCN}$.) To identify long-range forces at $\lambda > 10^{-6}$cm, one should compare the scattering cross section of UCN with the value of $4\pi b_{free\_int}^2$, where $b_{free\_int}$ is a scattering length measured by a neutron interferometer: $b_{free\_int} = b_{free\_nucl} + b_{long\_range}$. Probably the method using the scattering length measured by a neutron interferometer will be more insensitive (by a few centimeters) in the area of $\lambda$ than the coherence length of neutron beam in measurements by interferometers.

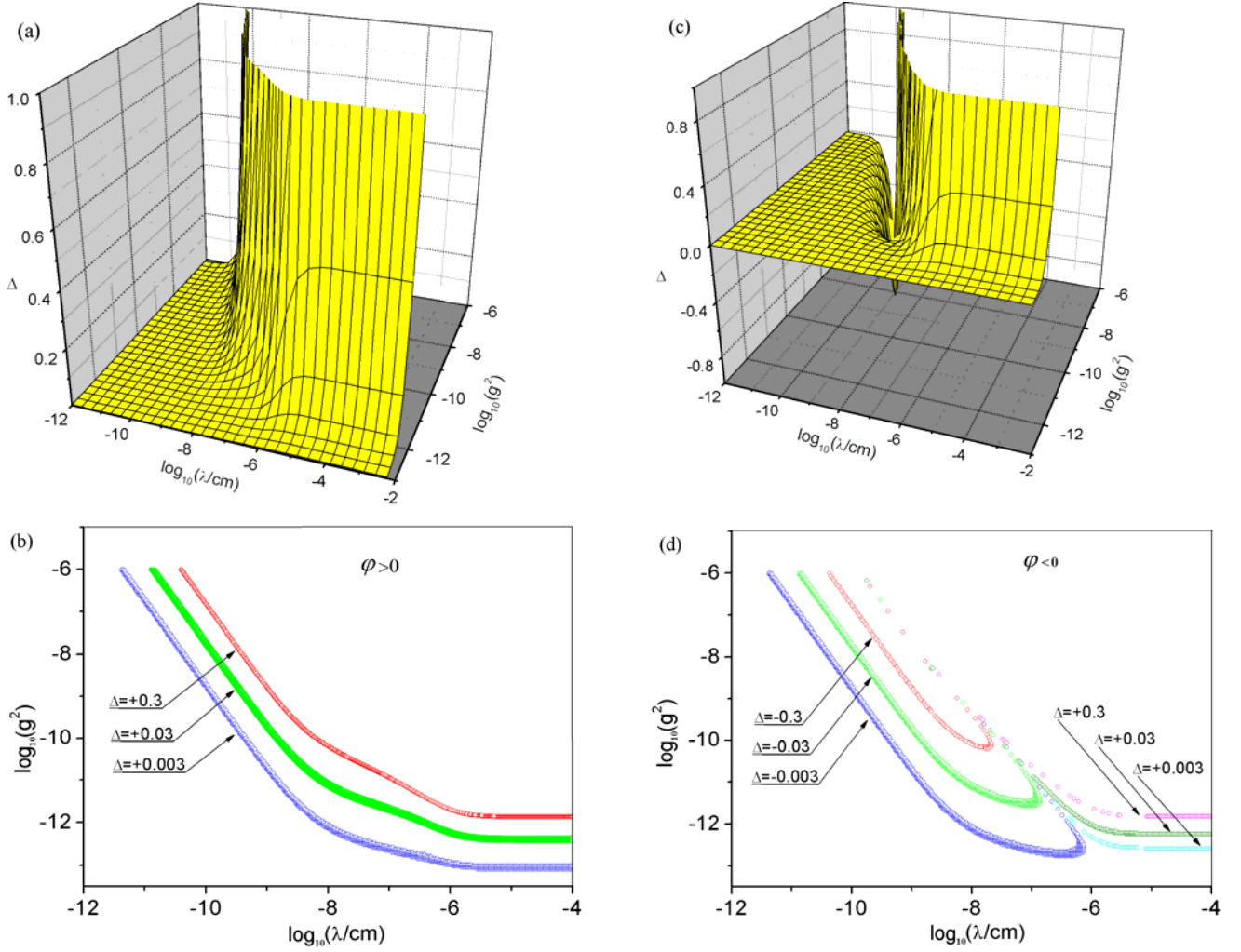

FIG. 3. Dependence of an expected experimental effect $\Delta_1^{(P\tau)-nucl}$ on parameters $g^2$ and $\lambda$ for He. (a) the case of a repulsive potential, i.e. an exchange by a vector boson. (b) correlation between parameters $g^2$ and $\lambda$ for a repulsive potential if the effect $\Delta_1^{(P\tau)-nucl}$ is equal to 0.3, 0.03 and 0.003. (c) the case of an attractive potential, i.e. an exchange by a scalar boson. (d) correlation between parameters $g^2$ and $\lambda$ for an attractive potential if the effect $\Delta_1^{(P\tau)-nucl}$ is equal to $\pm 0.3$, $\pm 0.03$ and $\pm 0.003$.

Recently we realized the preliminary measurements of $(P\tau)$ value for He with accuracy about 1% $\left( P\tau = (427 \pm 4 \text{mbar} \times \text{s}) \right)$. Results of our measurements are in agreement with results of the similar measurements published in work [25] $\left( P\tau = (467 \pm 33)\text{mbar} \times \text{s} \right)$.

Using the obtained value $(P\tau)$ we can be expressed the total cross section from formula $\sigma_{scat}^{He} + \sigma_{He\_capt}^0 V_{2200}/\bar{V}_{He} = (n_{He}\bar{V}_{He}\tau_{stor}^{gas})^{-1} = (P\tau_{stor}^{gas} \cdot 2.687 \cdot 10^{16} \cdot 273/293 \cdot \bar{V}_{He})^{-1}$, where 2.687 $10^{16}$ is the number of helium atoms in 1 cm$^3$ at pressure 1 mbar and temperature $T = 273K$, $\bar{V}_{He}$ is an average velocity of helium atoms at room temperature (293K) $\bar{V}_{He} = 1.240 \cdot 10^5 \text{cm} \times \text{s}^{-1}$; $P\tau = (427 \pm 4)\text{mbar} \times \text{s}$. Then $\sigma_{scat}^{He} + \sigma_{He\_capt}^0 V_{2200}/\bar{V}_{He} = (0.753 \pm 0.006) \times 10^{-24} \text{cm}^2$. The capture cross section of natural He because of admixture of He$^3$ is equal to $0.0075 \times 10^{-24}$ cm$^2$ for velocity 2200m/s, correspondingly the capture cross section for the average velocity $1.240 \cdot 10^5 \text{cm} \times \text{s}^{-1}$ is equal to $0.0133 \times 10^{-24} \text{cm}^2$. Then scattering cross section $\sigma_{scat}^{He}(\exp.P\tau) = (0.740 \pm 0.006) \times 10^{-24} \text{cm}^2$.

The nuclear scattering cross section $\sigma_{free\_nucl}^{He}$ measured by the transmission of neutron with energy from 0.19 eV up to 6.19 eV through volume with He gas [26] gives value $0.773 \pm 0.009 \times 10^{-24} \text{cm}^2$. The measurements carried out before [27] give value $0.73 \pm 0.05 \times 10^{-24}$ cm$^2$. In the tables of the neutron cross sections data [28] the recommended average value $0.76 \pm 0.01 \times 10^{-24}$ cm$^2$ is presented. Using this value we obtain the following $\Delta_1^{(P\tau)-nucl}$ value:

$$\Delta_{1,He}^{(P\tau)-nucl} = \left[ \frac{\sigma_{scat}^{He}(\exp.P\tau) e^{\frac{E_{UCN}^{trap}}{kT} \frac{(M+1)^2}{4M}}}{\sigma_{free\_nucl}^{He}} - 1 \right] = -0.026 \pm 0.015. \quad (1.7\,\sigma)$$

We do not see any real effect and for simplicity we can estimate upper limit for $\Delta_{1,He}^{(P\tau)-nucl}$ as +0.03 or as -0.03 at the confidence level of 95% ($2\sigma$). In Fig. 4 the restriction area ($g^2$, $\lambda$) at the confidence level of 95% is shown for the case of an attractive potential and for repulsive potential.

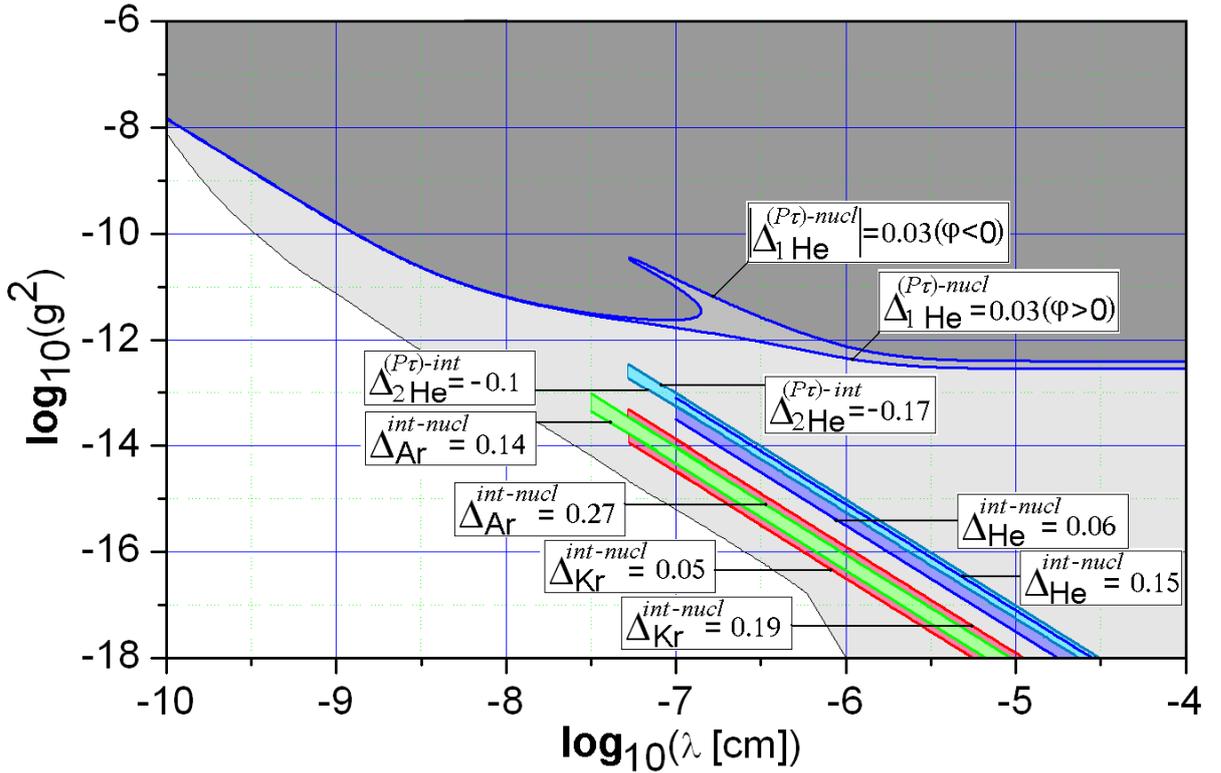

FIG. 4. The compactly shaded area corresponds to constrains for $g^2$ and $\lambda$ from our preliminary measurements of $(P\tau)$ value for He $(P\tau = (427 \pm 4 \text{mbàr} \times \text{s}))$ and the nuclear scattering cross section $\sigma_{free-nucl}^{He}$ ($0.76 \pm 0.01 \times 10^{-24}$ cm$^2$) measured by the neutron transmission [26, 27, 28]. The slightly shaded area corresponds to constrains for $g^2$ and $\lambda$ from works [9, 10]. The upper inclined area $\Delta_{2,He}^{(P\tau)-int}$ corresponds to values $g^2$, $\lambda$ from comparison of $(P\tau)$ method for He $(P\tau = (427 \pm 4 \text{mbàr} \times \text{s}))$ and the interferometer method [29] for He. The others inclined areas $\Delta_{He}^{int-nucl}$, $\Delta_{Ar}^{int-nucl}$, $\Delta_{Kr}^{int-nucl}$ corresponds to values $g^2$, $\lambda$ from comparison of the scattering cross sections measured by the transmission method [26] and by means interferometer [29] for He, Ar, and Kr. For all cases: $\Delta_{2,He}^{(P\tau)-int}$, $\Delta_{He}^{int-nucl}$, $\Delta_{Ar}^{int-nucl}$, $\Delta_{Kr}^{int-nucl}$ the solution exist only for repulsive potential.

Now it is necessary to compare the nuclear scattering cross section measured by means $(P\tau)$ - method and the coherent scattering cross section measured by means of interferometers. The measurements of scattering lengths by means of interferometers ($b_{free\_int}$) should also include the scattering length of long-range interaction ($b_{free\_int} = b_{free\_nucl} + b_{long\_range}$) since the scattering length is measured at zero scattering angle, i.e. $4\pi b_{free\_int}^2 = 4\pi(b_{free\_nucl} + b_{long\_range})^2$. Therefore we can obtain the following formula:

$$\Delta_2^{(P\tau)-int} = \left[ \frac{(\tau_{stor}^{gas} n_A \overline{V}_A)^{-1} - \sigma_{capt}^0 V_{2200}/\overline{V}_A}{4\pi b_{free\_int}^2} e^{\frac{E_{UCN}^{trap}}{kT}\frac{(M+1)^2}{4M}} - 1 \right] =$$

$$\pm g_\pm^2 M \left( \frac{(M+1)}{8\pi b_{free\_int}} \left( \frac{\hbar c}{kT} \right) e^{z(\lambda)} E_1(z(\lambda)) - \left( \frac{M}{M+1} \right) \frac{\lambda^2}{\pi b_{free\_int}} \frac{m_n c}{\hbar} \right) +$$

$$+ (g_\pm^2 M)^2 \left( \frac{(M+1)^2}{4(8\pi b_{free\_int})^2} \left( \frac{\hbar c}{kT} \right)^2 \frac{e^{z(\lambda)} E_2(z(\lambda))}{z} + \left( \frac{M}{M+1} \right)^2 \left( \frac{m_n c \lambda^2}{2\pi \hbar b_{free\_int}} \right)^2 - \right.$$

$$\left. - \left( \frac{m_n c M \lambda^2}{16\pi^2 \hbar b_{free\_int}^2} \right) \left( \frac{\hbar c}{kT} \right) e^{z(\lambda)} E_1(z(\lambda)) \right) \approx \left[ 1 \mp \frac{1}{2} J_0(g_\pm^2, \lambda) \right]^2 - 1, \qquad (17)$$

where $J_0(g_\pm^2, \lambda) = \frac{g_\pm^2 M^2}{\pi(M+1)} \frac{\lambda^2}{b_{free\_int} \lambdabar_C}$, $\lambdabar_C$ is the Compton neutron wave length $\lambdabar_C = \frac{\hbar}{m_n c}$.

The analysis using the simplified formula (17) for $\Delta_2^{(P\tau)-int}$ and for $\lambda > 10^{-7}$ cm is also shown in Fig. 4. The value $b_{bound\_int}$ was taken from work [30] ($b_{bound\_int,He} = 3.26 \pm 0.03$ fm). Since in our case one discusses scattering with a free nucleus, we should recalculate the scattering length for bound He nucleus in regard to the scattering length with a free nucleus of helium using the equation: $b_{free\_nucl,He} = b_{bound\_nucl,He} M_{He}/(M_{He} + 1) = 0.2608(24) \times 10^{-12}$ cm. Accordingly, the cross sections with a free nucleus, calculated from the scattering length with a free nucleus will be: $\sigma_0^{He} = 4\pi b_{free\_nucl,He}^2 = 0.855(16) \times 10^{-24}$ cm$^2$. At the same time the scattering cross section $\sigma_{scat}^{He}(\exp.P\tau) = (0.740 \pm 0.006) \times 10^{-24}$ cm$^2$. Then

$$\Delta_{2,He}^{(P\tau)-int} = \left[ \frac{\sigma_{scat}^{He}(\exp.P\tau) e^{\frac{E_{UCN}^{trap}}{kT}\frac{(M+1)^2}{4M}}}{4\pi b_{free\_int,He}^2} - 1 \right] = -0.134 \pm 0.018 \quad (7.5\ \sigma)$$

At the confidence level of 95% ($2\sigma$) the effect lies in the range (-0.17,-0.1). The area ($g^2, \lambda$) is shown in Fig.4 at the confidence level of 95% for the case of a repulsive potential. There is no solution for an attractive potential. One can see that the determined area of values $g^2$ and $\lambda$ is excluded by analysis results [9, 10].

It would be very interesting to have now precise $(P\tau)$ data for the heavy atoms. But it is not measured yet therefore we will use the scattering cross sections measured by the transmission of the collimated neutron beam with the different energy (from 0.07 eV up to 6 eV ) though the sample with the studied gas [26]. We can compare these cross sections with the cross sections measured by means interferometer [29] for the same gases. Making the comparison we have to take into account the correction for incoherent scattering cross section and we have to recalculate the bound cross section to the scattering cross section on the free atom. The results of the comparison are shown in Table 1. One can see that the values measured by the transmission method, except for a neon case, it is less than the similar values obtained from a neutron interferometer data.

TABLE I. The coherent cross sections $\sigma^*_{coh\_free}$ measured by the transmission method and coherent cross sections $\sigma^{**}_{coh\_free}$ measured by the neutron interferometer.

| Gas | Transmission method (Ref[26]) $\sigma^*_{coh\_free} \left[10^{-24} cm^2\right]$ | Neutron interferometer (Ref[29]) $\sigma^{**}_{coh\_free} \left[10^{-24} cm^2\right]$ | $\Delta\sigma = \sigma^*_{coh\_free} - \sigma^{**}_{coh\_free}$ |
|---|---|---|---|
| He | $0.773 \pm 0.009$ | $0.855 \pm 0.016$ | $-0.082 \pm 0.018 (4.5\sigma)$ |
| Ne | $2.42 \pm 0.03$ | $2.44 \pm 0.04$ | $-0.02 \pm 0.05 (0.4\sigma)$ |
| Ar | $0.424 \pm 0.008$ | $0.51 \pm 0.01$ | $-0.086 \pm 0.012 (7\sigma)$ |
| Kr | $6.19 \pm 0.17$ | $6.94 \pm 0.11$ | $-0.75 \pm 0.20 (4\sigma)$ |

The analysis for effect of the long-range forces can be done by means the following formula.

$$\Delta^{int-nucl} = \frac{\sigma^{**}_{coh\_free}}{\sigma^*_{coh\_free}} - 1 = \frac{1}{4}\left(\frac{g^2_\pm M^2}{\pi(M+1)}\right)^2 \left(\frac{\lambda^2}{b_{free\_nucl}\lambda_C}\right)^2 \pm \frac{g^2_\pm M^2}{\pi(M+1)} \frac{\lambda^2}{b_{free\_nucl}\lambda_C}. \quad (18)$$

Where the coherent cross sections $\sigma^*_{coh\_free}$ measured by the transmission method is equal $\sigma^*_{coh\_free} = \sigma_{scat\_free} - \sigma_{inc\_free}$. The total scattering cross section $\sigma_{scat\_free}$ was taken from Ref.[26]. Incoherent scattering cross section $\sigma_{inc\_free} = \left(M/(M+1)\right)^2 \sigma_{inc}$, the bound incoherent cross section $\sigma_{inc}$ is taken from Ref.[31]. $\sigma^{**}_{coh\_free} = 4\pi\left(M/(M+1)\right)^2 b_c^2$, where $b_c$ is taken from Ref.[29]. The value $b_{free\_nucl} = \sqrt{\sigma^*_{coh\_free}/4\pi}$.

The result of this analysis is presented in Fig. 4 at the confidence level of 95% ($2\sigma$). One can see that the determined area of values $g^2$ and $\lambda$ is excluded by analysis results [9, 10].

The presented discrepancy is likely to be due to a systematic experimental error. To clarify the problem new measurements are to be made for $(P\tau)$-data but it is also important to test the measurement of scattering lengths by means of interferometer. At last, if there is some chance that the observed difference is caused by long-range forces it could be tested by means the method of measuring the flux of above-barrier neutrons. This method is sensitive to the long-range forces also.

The method of measuring the flux of above-barrier neutrons will be considered in the next paragraph.

## 4. The method of measuring the flux of above-barrier neutrons

In case there is no gas in the trap the number of neutrons up scattered within an energy range from critical energy of foil $E_{UCN}^{foil}$ to that of the trap $E_{UCN}^{trap}$ is determined by UCN collision with the trap walls and is equal to

$$N_{low}(0) = \int_0^\infty N_0(0) e^{-t/\tau_{stor}^{total}(n_A=0)} \alpha v \, dt = N_0(0) \alpha v \tau_{stor}^{total}(n_A = 0), \qquad (19)$$

where $\alpha$ is probability per one collision that UCN will be up scattered within the range $E_{UCN}^{foil}$ - $E_{UCN}^{trap}$, $v$ is frequency of UCN collision with the trap walls, $\tau_{stor}^{total}(n_A = 0)$ is UCN storage time without gas, $N_0(0)$ is the number of UCN in the trap without gas at the moment of the beginning of lower up scattering effect measurements.

When gas at density $n_A$ is available the number of neutrons up scattered within the same energy range will be equal to:

$$N_{low}(n_A) = \int_0^\infty e^{-t/\tau_{stor}^{total}(n_A)} N_0(n_A) \left[ \alpha v + W_{at}^{low}(n_A) \right] dt = N_0(n_A) \left[ \alpha v + W_{at}^{low}(n_A) \right] \tau_{stor}^{total}(n_A), \qquad (20)$$

where $W_{at}^{low\_en}(n_A)$ is a probability that UCN will be up scattered within the range $E_{UCN}^{foil}$ - $E_{UCN}^{trap}$, $\tau_{stor}^{total}(n_A)$ is storage time of UCN in the trap with available gas at density $n_A$, $N_0(n_A)$ is the total number of UCN in the trap with gas at the moment of the beginning of measuring the lower energy up scattering effect.

Combining equations (19) and (20) one can obtain the following equation for subtraction of $W_{at}^{low}(n_A)$ from the experiment:

$$W_{at}^{low}(n_A) = \frac{N_{low}(n_A)}{N_0(n_A) \tau_{tot.st}^{up}(n_A)} - \frac{N_{low}(0)}{N_0(0) \tau_{tot.st}^{up}(0)}. \qquad (21)$$

Let us calculate $W_{at}^{low}(n_A)^{cal}$ taking into account the effect of long-range forces.

$$W_{at}^{low}(n_A) = \int_{\varepsilon_f}^{\varepsilon_t} d\sigma(\varepsilon) \int_{\varepsilon \frac{(M+1)^2}{4M}}^{\infty} \Phi(E) dE = 4\pi \frac{(M+1)^2}{M} n_A b_{free\_nucl}^2 \left( \frac{1}{2\pi m_n MkT} \right)^{1/2} \times$$

$$\times \int_{\varepsilon_f}^{\varepsilon_t} e^{-\varepsilon (M+1)^2 / 4M / kT} \left[ 1 \pm \frac{g_\pm^2 M^2}{\pi (M+1)} \frac{m_n c}{b_{free\_nucl} \hbar} \frac{\lambda^2}{2 m_n \varepsilon \lambda^2 / \hbar^2 + 1} + \right.$$

$$\left. + \left( g_\pm^2 M \right)^2 \frac{M^2}{(M+1)^2} \left( \frac{m_n c}{2\pi \hbar b_{free\_nucl}} \right)^2 \frac{\lambda^4}{\left( 2 m_n \lambda^2 \varepsilon / \hbar^2 + 1 \right)^2} \right] d\varepsilon . \qquad (22)$$

In calculating integral (22) we assume $\varepsilon_t$ and $\varepsilon_f$ to have the same order of magnitude, therefore a variable $\varepsilon$ is replaced by its average value of integration interval $\bar{\varepsilon} = (\varepsilon_t + \varepsilon_f)/2$. Now we can subtract contribution made by nuclear scattering and obtain the following equation for $g_\pm^2$ and $\lambda$.

$$W_{at}^{low}(n_A) \Big/ n_A \bar{V}_A \frac{(\varepsilon_t - \varepsilon_f)}{kT} 4\pi b_{free\_nucl}^2 \frac{(M+1)^2}{M} = \left[ 1 \pm \frac{1}{2} J_1 (g_\pm^2, \lambda) \right]^2 , \qquad (23)$$

where

$$J_1 (g_\pm^2, \lambda) = \frac{g_\pm^2 M^2}{\pi (M+1)} \frac{1}{b_{free\_nucl} \lambdabar_C} \frac{\lambda^2}{\lambda^2 / \bar{\lambdabar}_{UCN}^2 + 1} . \qquad (24)$$

Here, $\bar{\lambdabar}_{UCN}$ is de Broglie wave length ($\bar{\lambdabar}_{UCN}^2 = \hbar^2 / 2 m_n \bar{\varepsilon}$) of the neutron with kinetic energy $\bar{\varepsilon} = (\varepsilon_t + \varepsilon_f)/2$.

The left part of equation (23) implies the value $W_{at}^{low}(n_A)$ to be determined from the experiment according to equation (21). This equation contains the ratio of the counting rate of neutrons at energy above $E_{UCN}^{foil}$ to that of neutrons at energy below $E_{UCN}^{foil}$. The counting efficiency of neutrons depends on energy thus to achieve a very high accuracy of $W_{at}^{low}(n_A)$ is a certain problem. This problem can be solved using the method of relative measurements. For this purpose it is necessary to take the ratio of equations (23), for example, for $^{86}$Kr and for He.

$$\frac{\left[ W_{at}^{low}(n_A) \Big/ n_A \bar{V}_A \frac{(\varepsilon_t - \varepsilon_f)}{kT} 4\pi b_{free\_nucl}^2 \frac{(M+1)^2}{M} \right]_{Kr}}{\left[ W_{at}^{low}(n_A) \Big/ n_A \bar{V}_A \frac{(\varepsilon_t - \varepsilon_f)}{kT} 4\pi b_{free\_nucl}^2 \frac{(M+1)^2}{M} \right]_{He}} \approx$$

$$\approx \left\{ 1 \pm \left[ \frac{M_{Kr}^2}{(M+1)_{Kr} b_{free\_nucl}^{Kr}} - \frac{M_{He}^2}{(M+1)_{He} b_{free\_nucl}^{He}} \right] \frac{g_\pm^2}{2\pi \lambdabar_C} \frac{\lambda^2}{\lambda^2 / \bar{\lambdabar}_{UCN}^2 + 1} \right\}^2 \qquad (25)$$

In this ratio efficiencies of the detector are cancelled and the correction in the right part for He is much less than for $^{86}$Kr. From this formula one can see that for $\lambda > \bar{\lambdabar}_{UCN}$ the method sensitivity comes to saturation. In a similar way we can get the value $\Delta_3^{low\_energ-nucl}$:

$$\Delta_3^{low\_energ-nucl} = \frac{\left[ W_{at}^{low}(n_A) \Big/ n_A \bar{V}_A \frac{(\varepsilon_t - \varepsilon_f)}{kT} 4\pi b_{free\_nucl}^2 \frac{(M+1)^2}{M} \right]_{Kr}}{\left[ W_{at}^{low}(n_A) \Big/ n_A \bar{V}_A \frac{(\varepsilon_t - \varepsilon_f)}{kT} 4\pi b_{free\_nucl}^2 \frac{(M+1)^2}{M} \right]_{He}} - 1 . \qquad (26)$$

In Fig. 5 (a), (b) values of possible effect $\Delta_3^{low\_energ-nucl}$ are presented in the form of a surface for cases of positive and negative potentials $\varphi$.

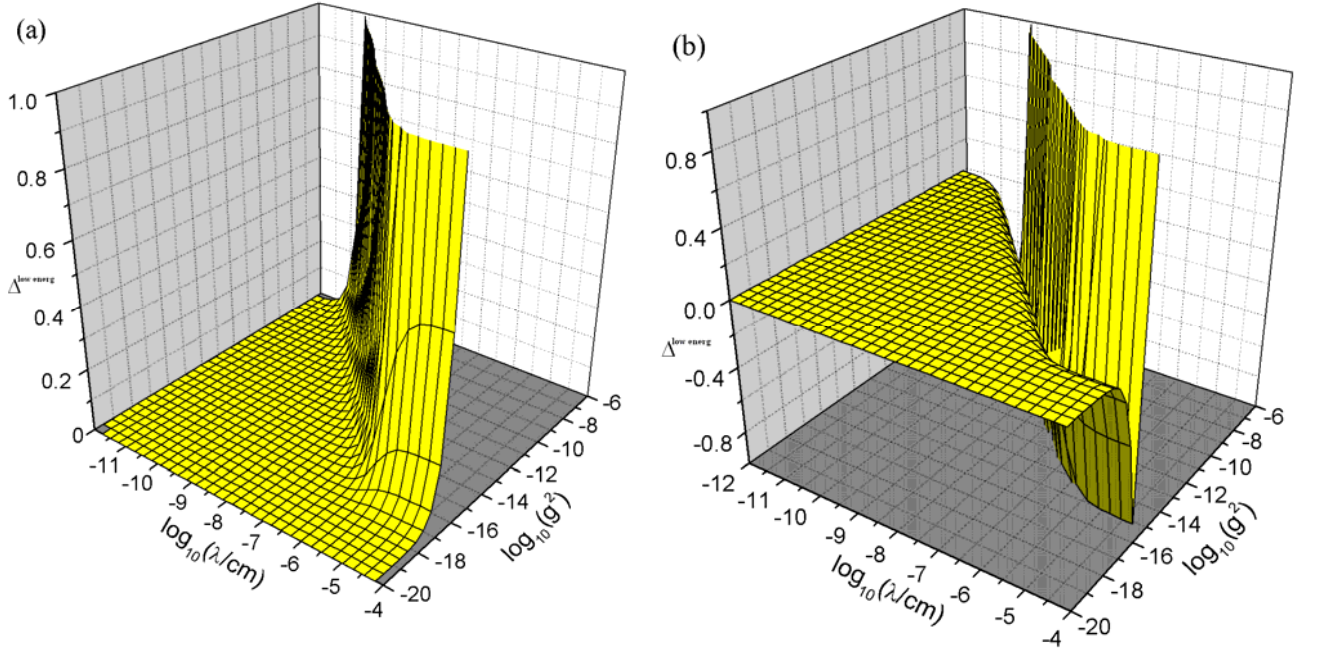

FIG. 5. Dependence of $\Delta_3^{low\_energ-nucl}$ on values $g^2$ and $\lambda$: (a) for $\varphi > 0$ and (b) for $\varphi < 0$.

Now we would like to make an analysis using the coherence neutron scattering length $b_{free\_int}$ derived from interferometer measurements. In this case we can obtain the following formula:

$$\frac{\left[W_{at}^{low}(n_A) \Big/ n_A \overline{V}_A \frac{(\varepsilon_t - \varepsilon_f)}{kT} 4\pi b_{free\_int}^2 \frac{(M+1)^2}{M}\right]_{Kr}}{\left[W_{at}^{low}(n_A) \Big/ n_A \overline{V}_A \frac{(\varepsilon_t - \varepsilon_f)}{kT} 4\pi b_{free\_int}^2 \frac{(M+1)^2}{M}\right]_{He}} \approx$$

$$\approx \left\{ 1 \mp \left[ \frac{M_{Kr}^2}{(M+1)_{Kr} b_{free\_int}^{Kr}} - \frac{M_{He}^2}{(M+1)_{He} b_{free\_int}^{He}} \right] \frac{g_\pm^2 \lambda^2}{2\pi \lambdabar_C} \left( \frac{\lambda^2}{\overline{\lambdabar}_{UCN}^2} \right) \frac{1}{\lambda^2 / \overline{\lambdabar}_{UCN}^2 + 1} \right\}^2 \quad (27)$$

Again we can introduce value $\Delta_4^{low\_energ-int}$:

$$\Delta_4^{low\_energ\_int} = \frac{\left[W_{at}^{low}(n_A) \Big/ n_A \overline{V}_A \frac{(\varepsilon_t - \varepsilon_f)}{kT} 4\pi b_{free\_int}^2 \frac{(M+1)^2}{M}\right]_{Kr}}{\left[W_{at}^{low}(n_A) \Big/ n_A \overline{V}_A \frac{(\varepsilon_t - \varepsilon_f)}{kT} 4\pi b_{free\_int}^2 \frac{(M+1)^2}{M}\right]_{He}} - 1 \quad (28)$$

In this case the method becomes sensitive for $\lambda > \overline{\lambdabar}_{UCN}$.

Now let us compare $(P\tau)$- method and that of above-barrier neutrons for two cases of analysis: with $b = b_{free\_nucl}$ and with $b = b_{free\_int}$. These results are shown in Fig. 6 (a,b,c,d). We have to regard that:

$\Delta_1^{(P\tau)-nucl}$ is $(P\tau)$- measurement of the scattering cross section in comparison with the nuclear scattering cross section measured at neutron energy about 1eV,

$\Delta_2^{(P\tau)-int}$ is $(P\tau)$- measurement of the scattering cross section in comparison with $4\pi b_{free\_int}^2$ measured by means of a neutron interferometer, $\Delta_3^{low\_energ-nucl}$ is the method of above-barrier neutrons in comparison with the nuclear scattering cross section measured at neutron energy about 1eV,

$\Delta_4^{low\_energ-int}$ is the method of above-barrier neutrons in comparison with $b_{free\_int}$ measured by means of a neutron interferometer.

If one assume that $\Delta$ is determined at accuracy about 1 % (for C.L. 95% upper limit of $\Delta$ is about +0.03 or – 0.03). The dependence of $g_\pm^2$ on $\lambda$ will look like, for example for krypton-86, as it is shown in Fig. 6 (a, b, c, d). The analysis is made for the case of $\varphi > 0$ (a, c) and for the case of $\varphi < 0$ (b, d). One can see that independence of $g_\pm^2$ from $\lambda$ is already observed within the area $10^{-8} – 10^{-4}$cm for $\Delta_1^{(P\tau)-nucl}$ but for $\Delta_3^{low\_energ-nucl}$ the independence of $g_\pm^2$ from $\lambda$ is within the area $10^{-6} – 10^{-4}$cm. For $\Delta_2^{(P\tau)-int}$ and $\Delta_4^{low\_energ-int}$ the independence of $g_\pm^2$ from $\lambda$ will take place within the area of $\lambda$ exceeding the length of neutron beam coherence in measurements with interferometers by a few centimeters.

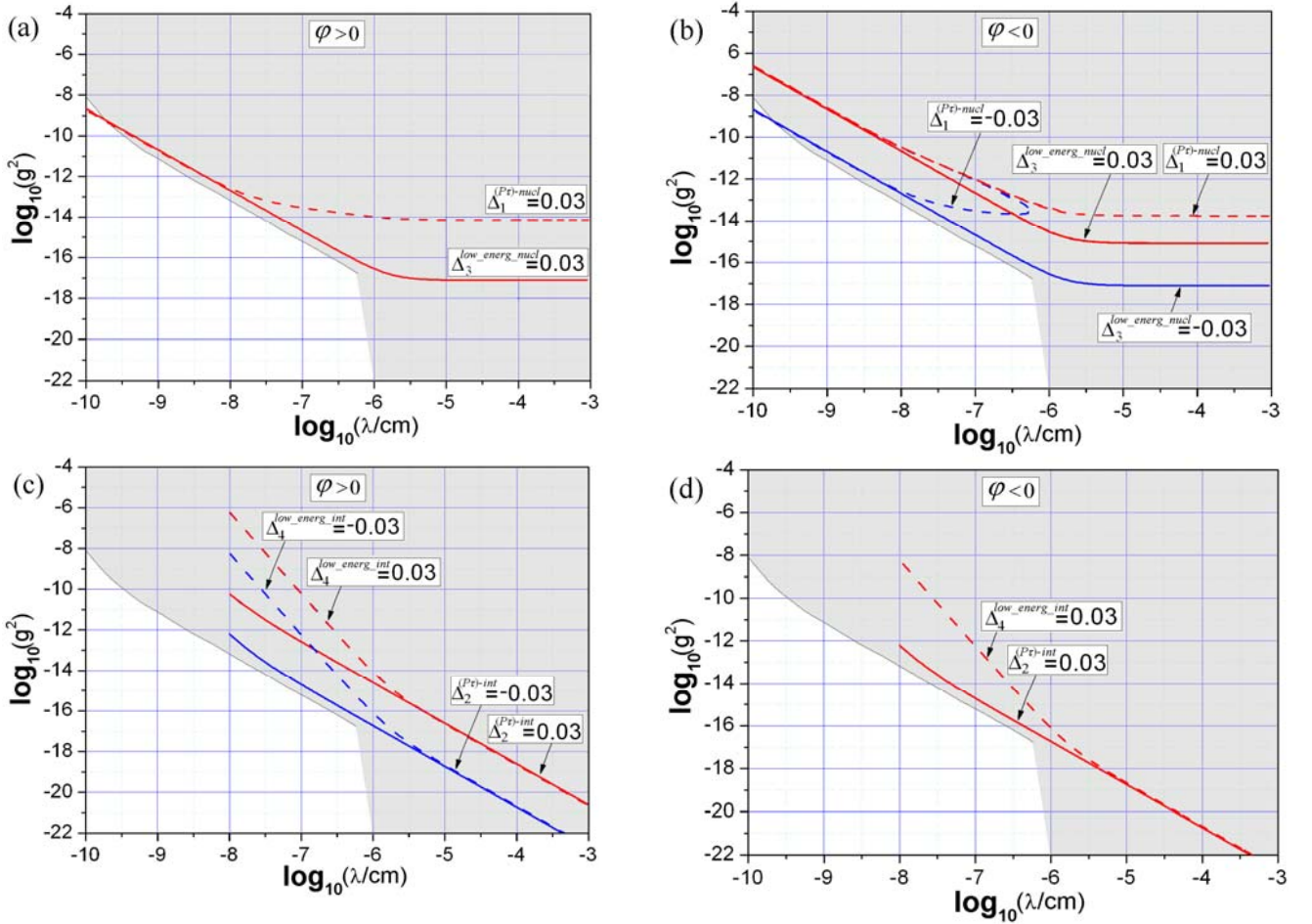

FIG. 6. The comparison of the constrains (95% C.L.) from $(P\tau)$ - method and method of above-barrier neutrons for two cases of analysis: with $b = b_{free\_nucl}$ and with $b = b_{free\_int}$ for the case of $^{86}Kr$. (a) $\varphi > 0$, $\Delta_1^{(P\tau)-nucl} = 0.03$, $\Delta_3^{low\_energ-nucl} = 0.03$; (b) $\varphi < 0$, $\Delta_1^{(P\tau)-nucl} = 0.03$ or -0.03, $\Delta_3^{low\_energ-nucl} = 0.03$ or -0.03; (c) $\varphi > 0$, $\Delta_2^{(P\tau)-int} = 0.03$ or -0.03, $\Delta_4^{low\_energ-int} = 0.03$ or -0.03; (d) $\varphi < 0$, $\Delta_2^{(P\tau)-int} = 0.03$, $\Delta_4^{low\_energ-int} = 0.03$. The painted area is excluded values $g^2$ and $\lambda$ from works [9, 10].

Now we would like to discuss some details of the experiment on above-barrier neutrons. It is necessary to take into account that values $N_{low\_en}$ and $N_0$ in equations (19-21) are different from the detector counting rate $N^*_{low\_en}$ and $N^*_0$. The probability of detection of a neutron is determined by the ratio of probability of neutron leakage to the detector $\tau^{-1}_{emp}$ to the total probability of UCN losses $\tau^{-1}_{emp} + \tau^{-1}_{stor}$, i.e. by the ratio $\dfrac{\tau^{-1}_{emp}}{\tau^{-1}_{emp} + \tau^{-1}_{stor}}$, where $\tau_{emp}$ is the time of UCN leaving the trap, and $\tau_{stor}$ is UCN storage time in the trap. Besides, it is necessary to take into account the efficiency of the detector ($D_{up}$) for corresponding energy range and also transmission foil factor ($T_{foil}$).

Thus, the number of neutrons $N_{low\_en}$, at energy above $E^{foil}_{UCN}$, but below $E^{trap}_{UCN}$, is connected with the counting rate of detector $N^*_{low\_en}$ by the following relation:

$$N_{low}(n_A) = \frac{N^*_{low}(n_A)}{d_{up}(n_A)}, \qquad (29)$$

where the factor $d_{up}(n_A) = T_{foil} D_{up} \tau^{up}_{stor}(n_A) / (\tau^{up}_{stor}(n_A) + \tau^{up}_{emp}(n_A))$.

The number of neutrons $N_0(n_A)$, whose energy is in the range from 0 to $E^{foil}_{UCN}$, is connected with the counting rate of detector by the following relation:

$$N_0(n_A) = \frac{N^*_0(n_A)}{d_{below}(n_A)}, \qquad (30)$$

where the factor $d_{below}(n_A) = D_{below} \tau^{below}_{stor}(n_A) / (\tau^{below}_{stor}(n_A) + \tau^{below}_{emp}(n_A))$.

The values $\tau^{below}_{stor}(n_A)$ and $\tau^{below}_{emp}(n_A)$ are measured at absorber position $h = E^{foil}_{UCN}/mg$ and with removed exit foil (6). The values $\tau^{up}_{stor}(n_A)$ and $\tau^{up}_{emp}(n_A)$ are measured at the top position of absorber when $h = E^{trap}_{UCN}/mg$ and with the entrance foil (1) at the neutron guide entrance. The factor of the foil transmission $T_{foil}$ is measured in the same case, comparing the measurements with and without the exit foil (6). As it was already mentioned the values $D_{up}$ and $D_{below}$ for measurements with He and with $^{86}$Kr are cancelled in the relation (25). Thus, it is obviously possible to reach accuracy about 1 % in determination of the left part of equation (25).

## 5. Conclusion

According to Fig. 6 we can conclude that the method of above-barrier neutrons is more sensitive with respect to ($P\tau$) method within the range of large $\lambda$ values. It is concerned with possibility of direct detection of neutrons scattered due to long-range forces. At the same time the analysis with $b = b_{free\_int}$ allows to extent considerably the range of studied $\lambda$ values.

The detailed measurements of dependence of $P_A \tau^{gas}_{stor}$ value on gas pressure are necessary in order to study some possible effect of collective interaction of UCN with gas atoms.

To summarize it is worth noticing that sensitivity of proposed methods is comparable with that of methods discussed in introduction [9] but in this case neutrons with extremely low energy are used. The effect of long-range interaction (if it exists) can be isolated directly by means of registration of lower energy up scattered neutrons. In the proposed methods we consider interaction of neutron with free atoms. It is important to study the long-range forces with radius exceeding the distance between atoms in materials.

This work was supported by the Russian Foundation for Basic Research (project nos. 08-02-01052a, 10-02-00217a, 10-02-00224a) and by the Federal Agency of Education of the Russian Federation (contract nos. P2427, P2500, P2540), by Ministry of Education and Science of the Russian Federation (contract nos. 02.740.11.0532, 14.740.11.0083).